# SCENAR$_{IOT}$CHECK: A Checklist-based Reading Technique for the Verification of IoT Scenarios

**Bruno Pedraça de Souza[1] and Guilherme Horta Travassos[1]**

[1]Department of Systems Engineering and Computer Science – Federal University of Rio de Janeiro – COPPE/UFRJ, Rio de Janeiro, RJ, Brazil

`{bpsouza, ght}@cos.ufrj.br`

***Abstract.*** *Software systems on the Internet of Things have driven the world into a new industrial revolution, bringing with it new features and concerns such as autonomy, continuous device connectivity, and interaction among systems, users, and things. Nevertheless, building these types of systems is still a problematic activity due to their specific features. Empirical studies show the lack of technologies to support the construction of IoT software systems, in which different software artifacts should be created to ensure their quality. Thus, software inspection has emerged as an alternative evidence-based method to support the quality assurance of artifacts produced during the software development cycle. However, there is no knowledge of inspection techniques applicable to IoT software systems. Therefore, this research presents SCENAR$_{IOT}$CHECK, a Checklist-based Reading Technique for the Verification of IoT Scenarios. The checklist has been evaluated with experimental studies. This research shows that the technique has good results regarding cost-efficiency, efficiency, and IoT software system development effectiveness.*

***Resumo.*** *Os sistemas de software em Internet das Coisas conduziram o mundo a uma nova revolução industrial, trazendo consigo novos recursos e preocupações, como autonomia, conectividade contínua de dispositivos e interação entre sistemas, usuários e coisas. No entanto, construir esses tipos de sistemas ainda é uma atividade difícil devido às suas características específicas. Estudos empíricos mostram a falta de tecnologias para apoiar a construção de sistemas de software de IoT, nos quais diferentes artefatos de software devem ser criados para garantir sua qualidade. Assim, a inspeção de software surgiu como um método alternativo baseado em evidências para apoiar a garantia de qualidade dos artefatos produzidos durante o ciclo de desenvolvimento do software. Todavia, não há conhecimento das técnicas de inspeção aplicáveis aos artefatos criados a partir dos sistemas de software IoT. Portanto, esta pesquisa apresenta SCENAR$_{IOT}$CHECK, uma técnica de leitura baseada em checklist para verificação de cenários IoT. O checklist foi avaliado com estudos experimentais. Esta pesquisa mostra que a técnica tem resultados satisfatoriamente em relação ao custo-benefício, eficiência e eficácia na construção de sistema de software de IoT.*

## 1. Introduction

Currently, citizens experience the 4$^{th}$ Industrial Revolution (or Industry 4.0) scenario, in which new interaction opportunities emerge among things, humans, machines, and systems (LIAO *et al.*, 2017; MOTTA, DE OLIVEIRA, and TRAVASSOS, 2018). One of the purposes of Industry 4.0 is to promote and improve the lives of end-users, making software systems and machines independent (LIAO *et al.*, 2017). It embraces new types of software systems, such as those involving the Internet of Things (or IoT) paradigm, context awareness capability, ubiquity characteristics, and cyber-physical properties (MOTTA, DE OLIVEIRA, and TRAVASSOS, 2018). All these software systems come up with specific features that should be carefully treated in the life cycle development (from requirements to deployment).

Especially for IoT software systems, these new characteristics adopt the autonomy to decide and act in all contexts, continuous connectivity among devices, smartness capacity, and others (MOTTA, DE OLIVEIRA, and TRAVASSOS, 2018; SOUZA, MOTTA and TRAVASSOS, 2019a). In general, the IoT paradigm is defined as composing software systems with uniquely addressable objects (*things*), equipped with identifying, sensing, processing, or actuation behaviors and processing capabilities that can communicate and cooperate to reach a specific goal. Besides, it bridges the real and virtual worlds through the internet connection (ATZORI, IERA, and MORABITO, 2010; MOTTA, SILVA, and TRAVASSOS, 2019).

Given this scenario, the engineering of IoT software systems can be challenging and demands new or adjusted software technologies to construct and guarantee their quality (ZAMBONELLI, 2017). In this context, we identified several emergent technologies to elicit and specify IoT software systems requirements available in the technical literature (REGIO, 2018; SILVA, 2019; FERRARIS and FERNANDEZ-GAGO, 2019; SILVA, GONÇALVES, and TRAVASSOS, 2020). Nevertheless, we could not find any specific technology that applies a verification technique to support the quality of the requirements artifacts. Thus, as a research opportunity, we proposed a new technology of software verification. Consequently, we contributed to evolving software Verification and Validation (V&V) area (BOEHM, 1984).

Software Engineering (SE) commonly incorporates activities of Quality Assurance (QA) because it contributes to ensuring that the constructed artifacts reach the established quality levels. Several techniques can be applied to the artifacts produced during the life cycle (from requirements to testing) with different V&V approaches (BOEHM, 1984). For instance, Software Reviews or Software Testing. Software Reviews are static verification techniques such as auditory, walkthroughs, and software inspections[1] (FAGAN, 1986).

In this work, we focused on the software inspection for IoT software systems. Our goal is to contribute to the evolution of the SCENAR$_{IOT}$ Technique and, consequently, to the Software V&V area. Thus, the main research question of this work is: *How to inspect IoT software systems requirements to verify whether the components and properties specified with SCENAR$_{IOT}$ were captured/described properly in the scenarios?*

Therefore, we propose a new checklist-based inspection technique called SCENAR$_{IOT}$CHECK (SOUZA, 2020). It aims to support IoT scenario descriptions produced by SCENAR$_{IOT}$ Technique (SILVA, 2019; SILVA and TRAVASSOS, 2020). SCENAR$_{IOT}$ Technique is a scenario-based requirements specification technique specific for IoT software systems (SILVA, 2019; SILVA and TRAVASSOS, 2020). It is worth highlighting that the SCENAR$_{IOT}$CHECK consists of a checklist that supports software engineers to uncover possible requirements problems in IoT scenarios descriptions.

---

[1] Software inspection is a static verification technique; it has risen as a formal and rigorous technique to assess artefacts produced from the software development cycle (FAGAN, 1986).

This paper is organized as follows, reflecting the research context, the main objective, and the research question. First, the methodology conducted in this research is described in Section 2; Section 3 presents the design of the SCENAR$_{IOT}$CHECK Technique, which is grounded on the SCENAR$_{IOT}$ Technique and IoT Facets. Section 4 presents the two experimental studies regarding SCENAR$_{IOT}$ and SCENAR$_{IOT}$CHECK Techniques. Next, Section 5 describes the threats to the validity of this research. Finally, Section 6 concludes by presenting the main contributes, future work, and final remarks.

## 2. Methodology

The methodology adopted in this work is based on evidence. It consists of carrying out a series of experimental studies to assess and improve the technology to be built (SHULL, CARVER, and TRAVASSOS, 2001). Figure 1 describes the methodology, divided into four steps to propose and evaluate the SCENAR$_{IOT}$CHECK Technique.

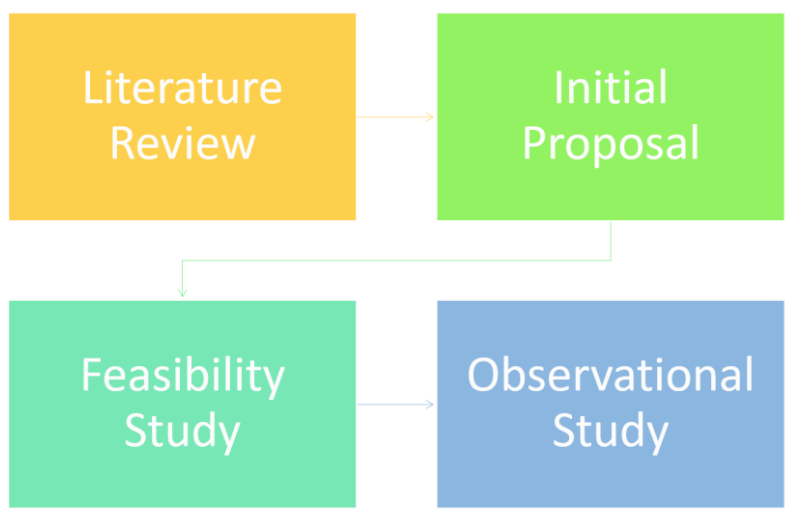

**Figure 1. Methodology Adopted in this Research.**

**Literature Review** aims at searching for primary or secondary studies to obtain the main requirements inspection techniques in the context of conventional (web-based and mobile apps) and/or contemporary software systems (IoT, ubiquitous systems, among others). Besides, we looked for secondary studies regarding IoT software systems characteristics, definitions, and their application.

**Initial Proposal** defines a set of questions for the first version of the SCENAR$_{IOT}$CHECK Technique according to SCENAR$_{IOT}$ Technique (SILVA, 2019) and IoT Facets (MOTTA, DE OLIVEIRA, and TRAVASSOS, 2019b).

**Feasibility Study** to assess whether the first version of SCENAR$_{IOT}$CHECK Technique achieves its objective of detecting defects in the scenario's description produced by the SCENAR$_{IOT}$ specification technique. So, the research question defined in this study was: *"Does the SCENAR$_{IOT}$CHECK Technique support the identification of defects in scenario descriptions for IoT software systems?"*

The purpose of the **Observational Study** was to evaluate and improve the second version of SCENAR$_{IOT}$CHECK concerning the comprehension of its application. In addition, to assessing the efficiency and effectiveness of the SCENAR$_{IOT}$CHECK Technique. In this study, the SCENAR$_{IOT}$ and SCENAR$_{IOT}$CHECK Techniques were evaluated together. Hence, the research question defined in this study was: *"Do the steps of the process make sense?"*

## 3. SCENAR$_{IOT}$CHECK Inspection Technique

The SCENAR$_{IOT}$CHECK inspection technique was proposed to assure the quality of IoT software system requirements. It is based on a questionnaire (checklist format). Furthermore, this inspection technique comprises two other approaches, the SCENARIOT Technique (SILVA, 2019) and IoT Facets (MOTTA, MARÇAL, and TRAVASSOS, 2019), to promote a higher level of coverage in the inspection of components in the IoT-based scenarios artifacts. In this section, we present the second version of SCENAR$_{IOT}$CHECK. This last version was evaluated experimentally by an observational study with undergraduate students (SOUZA *et al.,* 2019b).

### 3.1. SCENAR$_{IOT}$ Specification Technique

SCENAR$_{IOT}$ is a scenario-based requirement specification technique that aims to support the description of specific requirements for IoT software systems. Such a technique comprises nine IoT interaction arrangements (IIA) - Figure 2 (SILVA, 2019; SILVA and TRAVASSOS, 2020). Each IIA tailors and evolves common scenarios technologies to support the specification of IoT-based scenarios considering their characteristics and interactions. Hence, they have an associated catalog and support the specification of a particular scenario. Besides, it is possible to combine one or more IIAs to produce a new scenario.

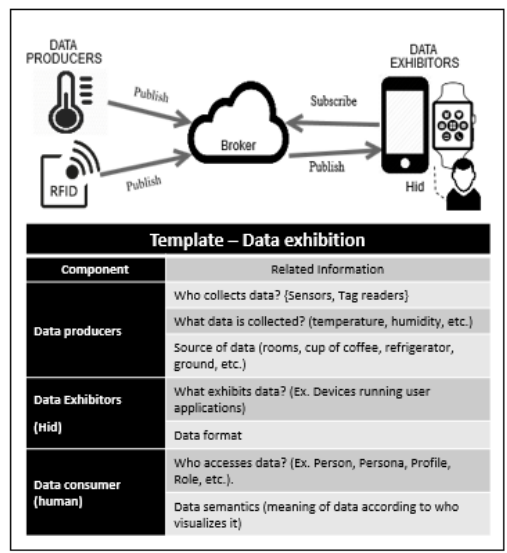

**Figure 2. Excerpt of an IIA and its catalog. Extracted from Silva and Travassos (2019b)**

### 3.2. IoT Facets

Non-functional features are present in traditional software systems (web and mobile). However, in the context of IoT-based systems, these non-functional proprieties are intensified due to their specific behavior to work together with sensors, identifiers, and devices. Motta, Silva, and Travassos (2018) carried out a structured literature review to identify the main challenges regarding non-functional proprieties to build IoT software systems. They found some IoT proprieties and organized them into six facets that should be addressed when engineering IoT-based software systems: Connectivity, Things, Behavior, Smartness, Interactivity, and Environment (MOTTA, SILVA, and TRAVASSOS, 2019a; MOTTA, DE OLIVEIRA, and TRAVASSOS, 2019b). Figure 3 shows how the authors represent the matrix involving the non-functional properties and the roles.

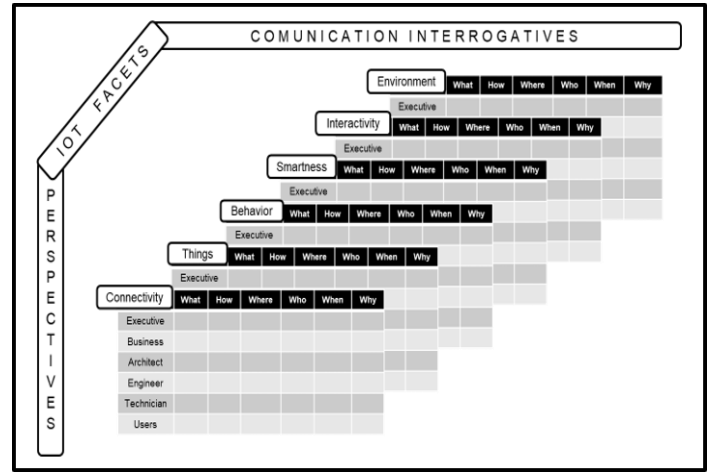

**Figure 3. IoT Facet organized in a Matrix. Extracted from Motta, De Oliveira and Travassos (2019b)**

### 3.3. SCENARIoTCHECK Inspection Technique

Initially, we identified and mapped the IoT facets within IIAs, looking to cover all the essential information regarding the IoT characteristics that should be inspected. Figure 4 shows an excerpt of the mapping between IoT facets within IIAs (SOUZA, 2020). Colors highlight the mapping between IoT facets within IIAs. For instance, green represents the devices, hardware, and things that need to be captured by the IIAs. An example of a question for this part is: "*Is it possible to identify who or what collects the data? (Sensors, QR code readers, people)*". The red color shows the type of connection that a software system must address. An example of a question about the connection is: "*Is the type of communication technology that the system adopts described in the scenarios? (Bluetooth, internet, among others)*".

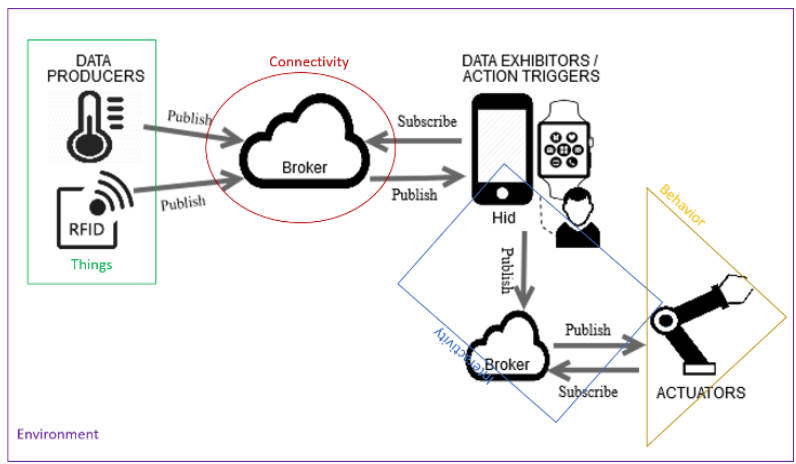

**Figure 4. Mapping between the IoT facets into the IIAs (SOUZA, 2020)**

After mapping the IoT facets within IIAs, we divided the SCENAR$_{IOT}$CHECK into two parts to organize the questions better. The first part (Figure 5) aims to verify if the essential and general information regarding IoT software systems, such as the main actors and roles used in the system, is the type of collected data.

| | | General Questions | | | Specific |
|---|---|---|---|---|---|
| General Questions | 01 | Has the overall application domain been established? (Health, leisure, traffic) | 14 | Is there any sequence of actions in confused comprehension scenarios? | |
| | 02 | Is the specific purpose of the system correctly described? (Data visualization, visualization, decision making, and actuation only) | 15 | Are the actors described in the scenarios consistent with the actors described in the arrangements? (Things, software systems, users) | |
| | 03 | Is the type of data collected specified? (Temperature, humidity, pollution) | 16 | Are the scenarios related to the arrangements correctly? | |
| | 04 | Is it possible to identify who or what collects the data? (Sensors, QR code readers) | 17 | Do the scenarios seek to be accurate by presenting title and flows? (Presenting the purpose and actions of the system directly and explicitly) | |
| | 05 | Is it possible to identify who or what manages the data collected? (Administrator, decision-maker, users) | 18 | Are adverbs that generate more than one possibility of interpretation in scenarios avoided? (Probably, possibly, supposedly) | |
| | 06 | Is it possible to identify who or what accesses the data collected? (Things, software systems, users) | 19 | Are condition terms (like "if", "go to", "while") used correctly? | |
| | 07 | Is the user interface device that displays the data described? (dashboard, smartphone, tablet) | 20 | When you use words like "things," "data" in the scenario, do they have the same meaning in other parts of the same scenario? | |
| | 08 | Is it possible to identify who is viewing the data? (Things, software systems, users) | 21 | Is it possible to identify "things" described with a given function in the arrangements that represent another function in the scenarios? | |
| | 09 | Is it possible to identify the source from which the data is provided? (Chairs, table, automobiles, houses, buildings) | 22 | Are the alternative and/or exception flows described? | |
| | 10 | Are the roles involved in the system described? (Things, software systems, users) | 23 | Does the scenario specification identify the matching ID arrangement? (AII1, AII2, ..., AII9) | |
| | 11 | Is there any description of each actor in the specified scenarios? | | | |
| | 12 | Is it possible to identify the source of data provision? | | | |
| | 13 | Has each action within the scenario been described clearly and contains no extraneous information? | | | |

**Figure 5. The first part of SCENAR$_{IOT}$CHECK (SOUZA, 2020)**

The second part (IoT facets) of the SCENAR$_{IOT}$CHECK consists of checking specific questions related to non-functional properties, such as interactions among actors, the smartness of the systems, and the connectivity of the systems and devices. Each IoT facet has at least one question about it. Figure 6 presents the questionnaire.

| | | |
|---|---|---|
| Specific Questions | 24 | Is it possible to identify the specific context in which the system is embedded? (Smart room, smart greenhouse, autonomous vehicle, healthcare) |
| | 25 | Are the limitations of the environment described? (e.g., lack of connectivity structure, lack of hardware structure, inadequate infrastructure) |
| | 26 | Are the technologies associated with system objects described? (smartphones, smartwatches, wearables) |
| | 27 | Are the events that the system has identified? (e.g., on/off an object, sending data) |
| | 28 | What kind of communication technology does the system use in the scenarios? (Bluetooth, intranet, internet ...) |
| | 29 | Does the proposed communication technology meet the geographic/physical specifications of the system? (Large, medium or small scale) |
| | 30 | Is it possible to identify how the system will react according to changes in the environment? |
| | 31 | Are the interactions between the system and the environment represented in the scenarios? |
| | 32 | Is it possible to identify the interaction between actors? |

**Figure 6. The second part of SCENAR$_{IOT}$CHECK (SOUZA, 2020)**

In addition, to support both the SCENAR$_{IOT}$ technique and the software engineers in describing IoT scenarios in this work. Figure 7 shows an excerpt of the proposed template. The template works together with SCENAR$_{IOT}$ Technique since it gathers essential IoT software systems characteristics that should be specified.

| System Goal | [describe the purpose of this system] |
|---|---|
| System Domain | [describe the system domain, such as health, leisure, transit, and so on] |
| Actors | [describe system actors such as users; devices; software systems;] |
| Types of Collected Data | [describe types of data collected by sensors, such as temperature; humidity; time; luminosity; and so on] |
| Scenario ID | SC[id]    Title    [scenario title] |
| Arrangement | [IIA-01, ..., IIA-09] and name of arrangement |
| Actors | *Users:* [describe users such as: end user, animals...] <br> *Devices:* [describe things with their sensors, actuators, wearables...] <br> *Software systems:* [describe software systems such as web systems, mobile app, and so on] |
| Steps (All system steps must be described in detail) | **Interaction Sequence** <br> [MAIN FLOW - Describe the stages of the scenario using the actors described above and their respective interactions in the arrangement. Data collection and processing must also be contemplated. Remember, scenarios need to be objective and clearly understood]. <br> [ALTERNATIVE FLOW - describe the alternative flows in the scenario]. <br> [EXCEPTION FLOW - Describe the exception streams in the scenario]. |

**Figure 7. Excerpt of the scenario template (SOUZA, 2020)**

## 4. Experimental Studies

This section presents the two experimental studies regarding the evaluation of the SCENAR$_{IOT}$CHECK Technique. The first one is a feasibility study that aims to verify whether the technique detects defects in IoT scenarios description (SOUZA, MOTTA, and TRAVASSOS, 2019a). The second study is classified as observational, where the purpose was to observe how the inspectors applied the SCENAR$_{IOT}$CHECK Technique (SOUZA *et al.,* 2019b).

### 4.1. Feasibility Study

The main question of this feasibility study is: *"Does the SCENAR$_{IOT}$CHECK technique support the identification of defects in scenario descriptions for IoT software systems?"* Besides, this study is organized by GQM (BASILI, CALDIERA, and ROMBACH, 1994) that aims to **analyze** the SCENAR$_{IOT}$CHECK technique **with the purpose of** characterizing **in relation to** its cost-efficiency, efficiency, and effectiveness **from the point of view** of software engineering researchers **in the context of** developers (represented by students of SE) evaluating an IoT scenario description artifact produced with SCENAR$_{IOT}$ specification technique.

*4.1.1. Planning*

The study was executed during a summer course at the Federal University of Amazonas in mid-January of 2019. The classes lasted four days, where we divided into two hours/day. The topics discussed during the course was about IoT software systems and their definitions and applications. Additionally, we presented the topics on requirements engineering, such as specification techniques based on scenarios and static software verification.

      Firstly, a pilot study was conducted to verify the quality of the artifacts used in this feasibility study. Three graduated participants cooperated with the pilot study. The

artifacts generated during the study were SCENAR$_{IOT}$ and SCENAR$_{IOT}$CHECK techniques, the defects records, the consent form, the characterization of participants, and the questionnaire to obtain the feedback' subjects.

After verifying the quality of the artifacts, 16 participants were invited to participate in this study. The participants were divided into four groups. The participants were undergraduate students in software engineering and information systems courses.

The study used distinct metrics. *Cost-efficiency* is the ratio between the number of real defects detected per group divided by the total time spent in the inspection in hours. The *Efficiency* represents the ratio between the number of real defects divided by the number of discrepancies detected per group. *Effectiveness* is the ratio between the number of defects detected per group divided by the number of known defects in the artifact. The total *Time* of inspection was collected in hours (H). The *Number of defects/discrepancies* was collected by the number of real defects/discrepancies reported by each group. The *Number of known defects*: the sum of all (independent) defects identified in the artifacts by different groups throughout the inspection sessions.

*4.1.2. Execution*

The study began when we presented the issues regarding IoT and requirements engineering. Then, we conducted training on the requirement specification and inspection in the context of web-based and IoT systems. After that, we introduced both SCENAR$_{IOT}$ and SCENAR$_{IOT}$CHECK Techniques since they work together. Each day of training had two hours of running.

Then, we presented the proposed problem from which the groups should describe the scenarios using the SCENAR$_{IOT}$ technique. The problem was about a "*smart greenhouse*." The idea regards the building of a system for a greenhouse using the IoT paradigm. Each group created two possible scenarios on the proposed problem, totalizing eight scenarios described.

After describing the scenarios, the groups should inspect the produced scenarios. It is worth highlight that each group inspected the scenarios created by another group. For example, group 1 inspected the scenarios produced by Group 2. Additionally, in the first round of inspection, the groups applied the *ad-hoc* technique. In the second round, the groups applied the SCENAR$_{IOT}$CHECK technique. Thus, each group inspected two scenarios description of two different groups. For instance, group 1 applied the *ad-hoc* technique to inspect the scenarios produced by group 2. In the second round of inspection, group 1 applied the SCENAR$_{IOT}$CHECK technique in the scenarios produced by group 3. However, group 1 could not inspect the scenarios produced by group 4 because we only performed two inspection trials using ad-hoc and SCENARIOTCHECK. The cross-checking of the scenarios that each group inspected is represented in Table 1.

**Table 1. Study design of feasibility study**

| Group | Trial one (*ad-hoc*) | Trial two (*SCENAR$_{IOT}$CHECK*) |
|---|---|---|
| G1 | Scenario produced by G2 | Scenario produced by G3 |
| G2 | Scenario produced by G1 | Scenario produced by G4 |
| G3 | Scenario produced by G4 | Scenario produced by G1 |
| G4 | Scenario produced by G3 | Scenario produced by G2 |

*4.1.3. Results*

The defects found in the scenarios were 45 defects using both *ad-hoc* (called AH) and SCENAR$_{IOT}$CHECK (called SC) Techniques. Additionally, we noticed seven common defects in 45. So, our final set was 38 known defects in the artifacts. In addition, we analyzed the effectiveness (EFC), cost-efficiency (CE), and efficiency (EFF) of each technique.

In the context of the SCENAR$_{IOT}$CHECK technique, the groups found 36 discrepancies. Then we removed the duplicates, totaling 33 real defects in the average time of 0.6 hours. Regarding the effectiveness, the average was 0.608, where the cost-efficiency was 14.771, and the efficiency was 0.875.

Regarding the *ad-hoc* technique, the groups found ten real defects in an average time of 0.924 hours. About the effectiveness, the average was 0.436, where the cost-efficiency was 2.781, and the efficiency was 0.276. A summary of the main results is presented in Table *2*.

**Table 2. Summary of Results**

| Technique | Detect Defects | Inspection Time (Hours) | CE (%) | EFF (%) | EFC (%) |
|---|---|---|---|---|---|
| AH | 10 | 0.924 | 2.781 | 0.436 | 0.276 |
| SC | 33 | 0.60 | 14.771 | 0.608 | 0.875 |
| Legend: CE: *cost-efficiency*; EFF: *efficiency*; and EFC: *effectiveness* | | | | | |

We did not apply the descriptive statistics because the data points were too limited. So, we only calculated the median of the metrics applied. Nevertheless, the results showed that the SCENAR$_{IOT}$CHECK technique could be feasible since its indicators were higher than the *ad-hoc* technique.

Regarding the qualitative data, we gave a questionnaire to the groups to collect data. Then, we briefly present the analysis. It is worth mentioning that we did not apply a sophisticated technique to code the answer due to the small sample.

Concerning the understanding of the SCENAR$_{IOT}$CHECK Technique, the participants reported that, in general, the checklist questions are clear and use a language easy to comprehend. The usefulness of the SCENAR$_{IOT}$CHECK Technique has been observed due to it assisted the subjects in detecting defects in the description of IoT-based scenarios. However, in general aspects of the SCENAR$_{IOT}$CHECK Technique, a few questions were unclear or sometimes redundant. Some excerpts of the participants' opinion follow:

"*The simplicity of the questions.*" - From G1.

"*The checklist makes the process easier.*" - From G4.

"*Yes, because it evaluates from the interaction aspect to the scenario infrastructure.*" - From G2.

"*Yes, because it identifies the part about the inspection of the elements.*" - From G3.

"*The technique has information (questions) that I consider the same. Ex: question 2 = question 14.*" - From G4.

"*Some questions may be clearer, such as questions 08 and 11.*" - From G1.

## 4.2. Observational Study

The first study allowed us to observe the feasibility of the SCENAR$_{IOT}$CHECK Technique and evolve it according to the participants' suggestions (SOUZA, MOTTA, and TRAVASSOS, 2019a). Thus, after evolving the technique, we carried out an observational study to observe its second version (SOUZA et al., 2019b). Therefore, the main question of this observational study is "*Do the steps of the process make sense?*".

A GQM (BASILI, CALDIERA, and ROMBACH, 1994) structure describes the study as to **analyze** the SCENARIOTCHECK Technique **with the purpose of** understanding **in relation to** how inspectors apply the SCENAR$_{IOT}$CHECK technique **from the point of view** of software engineers **in the context of** novice developers evaluating an IoT scenario description artifact.

*4.2.1. Planning*

This observational study applied a similar design, as previously described (SOUZA, MOTTA, and TRAVASSOS, 2019a), but grouping. We used the same artifacts presented in the feasibility study. We only added one artifact that was the proposed template (Figure 7). Besides, a pilot study was performed with six students of a software engineering course at the Federal University of Rio de Janeiro. After the pilot study, we validated and corrected the artifacts used in the observational study.

We carried out this observational study in June 2019 in the software requirements discipline in the software engineering course at the Federal University of Amazonas with ten participants. The topics discussed in the course were requirements elicitation and specification, requirements verification and validation, and their applications. Also, we introduced topics on the IoT paradigm and its application and characteristics. It worth stressing that the inspection process (SAUER *et al.*, 2000; KALINOWSKI and TRAVASSOS, 2004) and the defects taxonomy (SHULL, 1999) were the same used in the first feasibility study (Section 4.1).

The metrics used in this study were: *Cost-Efficiency, Efficiency,* and *Effectiveness*. In addition, the independent variables identified were *participants' experience* and *knowledge*. The dependent variables *are efficiency, effectiveness, time*, and the *number of defects*.

Different from the feasibility study, we carried out three inspection trials. First, we used the *ad-hoc* technique to carry out the first trial. For instance, participant 1 inspected the scenario generated by participant 2 by using the *ad-hoc technique*. Then, the second trial used the SCENARIOTCHECK Technique to verify the same scenario inspected with the *ad-hoc* technique. At last, the SCENAR$_{IOT}$CHECK technique was applied in the third inspection trial to review another scenario description. For example, participant 1 inspected the scenario generated by participant 3 by using the SCENAR$_{IOT}$CHECK technique. We cross-checked in all scenario inspections.

*4.2.2. Execution*

We started this observational study by soliciting the participants to sign the consent and characterization forms. After that, we introduced the issue that they should perform the requirements description and inspection. The proposed issue was related to "*Animal Monitoring using the IoT paradigm.*" The main idea was to develop a software system for animal monitoring, obtaining information about their health and distance from the farm.

To do that, the animals would use a set of sensors and things.

Regarding the training, we, firstly, presented the requirements elicitation and verification topics. Next, we requested that the participants did an activity on the topics. After that, IoT characteristics and applications were introduced. Finally, we presented the SCENAR$_{IOT}$ and SCENAR$_{IOT}$CHECK Techniques.

So, the participants should describe the scenarios regarding a solution for this software system by using the SCENAR$_{IOT}$ technique. Each subject generated one scenario description, totaling ten different IoT-based scenarios inspected by both *ad-hoc* (our oracle) and SCENAR$_{IOT}$CHECK Technique. The characterization of the participants can be accessed in Souza (2020).

*4.2.3. Results*

Considering the metrics established for this study and its objective, we cannot directly compare *ad-hoc* (called AH) and SCENAR$_{IOT}$CHECK (called SC) since we performed three inspection trials.

The sum of defects found by all participants of this study was 124 discrepancies in the ten IoT scenarios applying both inspection techniques. Sixteen discrepancies were found in common (duplicates). Then, we removed them, totaling 108 known defects in all artifacts. After the discrimination meeting, we classified real/known defects. So, the participants identified 94 known defects in all scenarios. As the objective of our research question is to observe how the participants applied the SC technique, we focused on the participant's performance and evolution regarding the use of the technique itself, not to compare AH and SC techniques.

Summarizing the main results found in Table 3, we separated by time (hours), *cost-efficiency (CE), efficiency (EFF), and effectiveness (EFC)*. We can observe that both second (SC) and third (SC) trial had better performance than the first (AH) one. These results reinforce those results found in the feasibility study (Subsection 4.1) that the SC is more efficient in detect defects than AH.

**Table 3. Average indicators**

| Trial | Tech | Detect Defects | Inspection Time (Hours) | CE (%) | EFF (%) | EFC (%) |
|---|---|---|---|---|---|---|
| 1 | AH | 15 | 0.62 | 0,371 | 0,127 | 2,307 |
| 2 | SC | 45 | 0.64 | 0,642 | 0,396 | 7,954 |
| 3 | SC | 34 | 0.60 | 0,543 | 0,318 | 6,373 |
| Legend: Tech: Technique; CE: *cost-efficiency*; EFF: *efficiency*; and EFC: *effectiveness* | | | | | | |

We analyzed the SC technique in three perspectives CE, EFF, and EFC. Thus, we could compare both results from the feasibility and observational studies concerning the SC technique for verifying its evolution. Figure 8 shows the SC technique evolution. When we made such a comparison, the feasibility study results seemed better than the observational ones. This behavior may be associated with some events:

(i) The second version of the SCENAR$_{IOT}$CHECK Technique has more questions than the first one.

(ii) The studies' design differs in the grouping. The feasibility study grouped the inspectors while the participants performed the inspection by themselves in

(iii) There was a template for describing the IoT scenarios in the observational study, so the participants must use it.
(iv) The problem domains regarding IoT were a little different in the studies. The participants were supposed to describe scenarios of "*smart greenhouse*" in the feasibility study. However, in the observational study, they described scenarios on "*smart farm*ing."
(v) The participants performed two inspection trials using the SC technique in the observational study, while they performed only one inspection trial using the SC technique in the feasibility study. So, the "learning factor" should be considered.

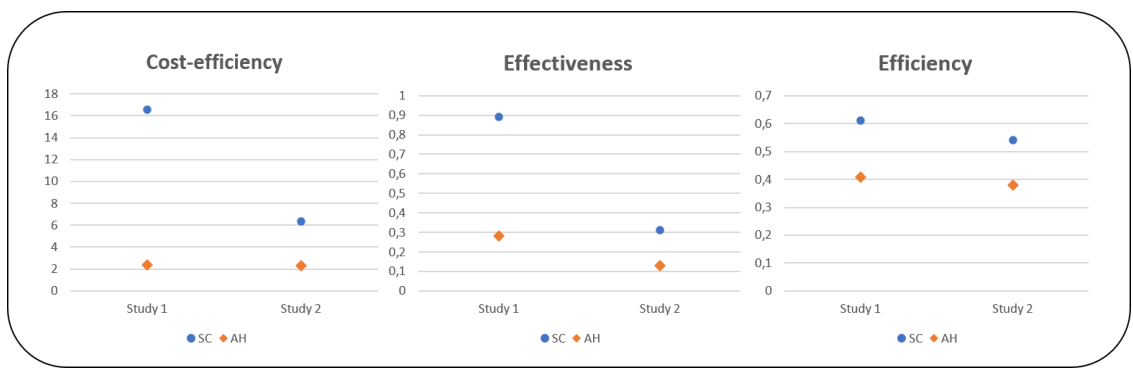

**Figure 8. The behavior/evolution of AH and SC in both studies**

## 5. Threats to Validity

In this section, the identified threats to the studies' validity. The Feasibility and Observational studies had a similar study design. Therefore, some threats can overlap each other.

Regarding **construct validity:** Despite providing the same problem for the exercise, we had no control of the artifacts produced during the execution of both studies. Besides, we could not guarantee that the scenarios produced were comparable on the number of defects and complexity. Concerning **external validity:** Despite the participants being undergraduate students, we mitigated this threat considering they have similar skills to novice software engineers (CARVER *et al.*, 2003). Additionally, we presented a real situation as the 'automation of a greenhouse' and 'smart farm' to them. **Conclusion validity:** In both studies, the sample size was small, limiting the generalization and conclusion of the obtained results. About **internal validity:** We prepared the training for *ad-hoc* and SCENAR$_{IOT}$CHECK techniques in both studies. In addition, we also applied a pilot study to observe if the equivalence on training.

## 6. Final Remarks, Contributions, and Future Work

Software verification and validation, especially software inspections, have brought many benefits to the software artifacts produced in the development cycle (from the requirements to the deployment). Over the years, several experimental studies regarding software inspections showed such benefits.

This dissertation presents the second version of the SCENAR$_{IOT}$CHECK

Technique, as well as its experimental evaluations. SCENAR$_{IOT}$CHECK Technique emerged in Industry 4.0 and IoT paradigm to maintain the quality assurance of the requirements artifacts generated by the SCENAR$_{IOT}$ Technique. IoT is an attractive emergent area of computer and Industry 4.0. Its main characteristics are the interaction among actors, devices, and software systems through a wireless network connection (internet).

To answer the main research question "*How to inspect IoT software systems requirements to verify whether the components and properties specified by SCENAR$_{IOT}$ were captured/described properly in the scenarios?"*, firstly, we performed a literature review looking for requirements inspection specific to IoT software systems since we have not found any inspection technique for IoT software systems. Then, we analyzed both SCENAR$_{IOT}$ and IoT facets characteristics to propose a new requirements inspection technique to support the QA of the IoT scenarios description. According to the results of both studies previously presented, we can assure that the SCENAR$_{IOT}$CHECK Technique is feasible to inspect IoT requirements artifacts.

We can highlight the following contributions of this work:

(i) Incorporating the software verification, requirements inspection in the context of the Fourth Industrial Revolution, presenting the main characteristics of these software systems.

(ii) The conception of a Checklist-based Reading Technique for the Verification of IoT Scenarios (SCENAR$_{IOT}$CHECK) in the context of IoT software systems. It was experimentally evaluated through two primary studies, showing its evolution, applicability in supporting the identification of defects, and assuring the quality of the SCENAR$_{IOT}$ scenarios (SOUZA, MOTTA, and TRAVASSOS, 2019a.; SOUZA *et al.,* 2019b).

(iii) To acquire and make explicit evidence from experimental studies in applying other requirements elicitation techniques in the context of the IoT software systems (SOUZA, MOTTA, and TRAVASSOS, 2019a; (SOUZA *et al.,* 2019b).

(iv) A template to support the description of IoT scenarios according to the SCENAR$_{IOT}$ catalogs.

We intend to carry out new experimental studies to include the SCENAR$_{IOT}$CHECK Technique in the software development cycle as future work. Also, we aim at applying a robust qualitative analysis technique to assess the qualitative data from both studies to evolve the SC technique to a new version. Next, we intend to observe how the SC technique fits in the real context of development. At last, we plan to build a computational infrastructure to support the process of both SCENAR$_{IOT}$ and SCENAR$_{IOT}$CHECK Techniques.

During this work, some papers related to the construction of the SCENAR$_{IOT}$CHECK Technique were published:

- **SOUZA, B. P.**; MOTTA, R. C.; TRAVASSOS, G. H. The first version of SCENAR$_{IOT}$CHECK: A Checklist for IoT-based Scenarios. In: XXXIII Brazilian Symposium on Software Engineering (SBES), Salvador, 2019, p. 219-223.

- **SOUZA, B. P.**; MOTTA, R. C.; COSTA, D. O.; TRAVASSOS, G. H. An IoT-based Scenario Description Inspection Technique. In: XVIII Brazilian Symposium on Software Quality (SBQS), 2019, Fortaleza. XVIII Brazilian Symposium on Software Quality (SBQS), 2019. v. 18.
- **SOUZA, B. P.;** MOTTA, R. C.; TRAVASSOS, G. H. Towards the Description and Representation of Smartness in IoT Scenarios Specification. In: XXXIII Brazilian Symposium on Software Engineering (SBES), 2019, Salvador. *(Best paper award –Insightful Ideas & Emerging Results Track).*
- SILVA, D. V.; **SOUZA, B. P**.; GONÇALVES, T. G.; TRAVASSOS, G. H. Uma Tecnologia para Apoiar a Engenharia de Requisitos de Sistemas de Software IoT. In: XXIII Congresso Ibero-Americano de Engenharia de Software (CIbSE), Curitiba, 2020. *(Best paper award – CIBSE – RET: Requirements Engineering Track).*

## Acknowledgment

We would like to thank all the studies' participants. We also thank Rebeca Motta and Taísa Guidini for their valuable reviews. This work was partially supported by CNPq and by Coordenação de Aperfeiçoamento de Pessoal de Nível Superior - Brasil (CAPES) - Finance Code 001. Prof. Travassos is a CNPq researcher.